\documentclass[letterpaper,tightenlines,nofootinbib,superscriptaddress]{revtex4}

\usepackage{amsmath}
\usepackage[bookmarks=false]{hyperref}
\usepackage{graphicx}
\usepackage{epstopdf}
\usepackage{color}

\newcommand{\abs}[1]{\lvert#1\rvert}

\renewcommand{\vec}[1]{\mathbf{#1}}

\newcommand{\df}{\mathrm{d}}

\newcommand{\mcdot}{\!\cdot\!}

\newcommand{\jetset}{{\{\mathrm{jets}\}}}

\newcommand{\cusp}{\!\mathrm{cusp}}
\newcommand{\ijpair}{\langle i,j \rangle}

\newcommand{\T}{\mathrm{T}} 

\newcommand{\as}{\alpha_s}

\newcommand{\ts}{\thinspace}
\newcommand{\dlog}[1]{\frac{\df#1}{\df\ts\ln\mu}}
\renewcommand{\ln}{\text{ln}\ts}
\newcommand{\new}{\nonumber\\}
\newcommand{\partset}{\{\mathrm{partons}\}}

\newcommand{\xone}{\sqrt{\frac{\tau}{z}}e^{Y}}
\newcommand{\xtwo}{\sqrt{\frac{\tau}{z}}e^{-Y}}
\newcommand{\fone}{f_1\left(\xone\right)}
\newcommand{\ftwo}{f_2\left(\xtwo\right)}
\newcommand{\zmin}{z_{\rm min}}
\newcommand{\order}[1]{{\cal O}(#1)}

\begin{document}


\title{On the effectiveness of threshold resummation away from hadronic endpoint}

\author{Christian W. Bauer}
\affiliation{Ernest Orlando Lawrence Berkeley National Laboratory,
University of California, Berkeley, CA 94720}

\author{Nicholas Daniel Dunn}
\affiliation{Ernest Orlando Lawrence Berkeley National Laboratory,
University of California, Berkeley, CA 94720}

\author{Andrew Hornig}
\affiliation{Department of Physics, Box 1560, University of Washington, Seattle, WA 98195
\vspace{2ex}}

\begin{abstract}
We parameterize the enhancement of threshold effects away from hadronic endpoint that arise due to the steeply falling nature of parton distribution functions, within the context of soft-collinear effective theory. This is accomplished in a process-independent way by directly linking the characteristic scale of soft and collinear radiation, $\lambda$, to the shape of the pdfs. This allows us quantify the power corrections to partonic threshold resummation as a function of the invariant mass and rapidity of the final state. In the context of SCET, being able to compute $\lambda$ in a process-independent manner allows us to determine the correct scale for threshold resummation after integration with the pdfs, without any additional procedure.
\end{abstract}

\maketitle

\section{Introduction}

There has been much effort to increase the accuracy and precision of predictions for observables at hadron colliders. The standard technique to improve the accuracy of calculations is to add fixed orders in perturbation theory. Many observables have been calculated to next-to-leading order (NLO) in QCD, and for some next-to-next-to-leading order (NNLO) has been achieved~\cite{Harlander:2002wh,Anastasiou:2002yz,Ravindran:2003um,Anastasiou:2003yy,Brein:2003wg,Anastasiou:2003ds,Anastasiou:2004xq,Anastasiou:2005qj,Anastasiou:2007mz,Grazzini:2008tf,Catani:2009sm}. Frequently, observables contain large ratios of scales, usually due to experimental cuts or the presence of several mass scales in the process. In such cases, one can increase the precision and accuracy of theoretical calculations by resumming the large logarithms of these ratios to all orders in perturbation theory. Standard techniques allow for the resummation of these logarithms at next-to-leading logarithmic order (NLL) and beyond  \cite{Sterman:1986aj,Catani:1989ne,Bonciani:2003nt}, while recent advances in effective theory methods  \cite{Bauer:2000ew,Bauer:2000yr,Bauer:2001ct,Bauer:2001yt} provide a systematically improvable alternative to resummation in hard-scattering processes \cite{Bauer:2002nz}.

In this paper, we will focus on so-called threshold logarithms. It is well known that, beyond leading order, partonic cross sections contain terms of the form
\begin{equation}
\label{eq:threshold_terms}
\alpha_s^n \left( \frac{\log^m(1-z)}{1-z}\right)_+\,,
\end{equation}
where $m \le 2n-1$ and  $z = \hat s_{\rm min} / \hat s$ is a measure of excess radiation in the process. These terms are due to collinear and soft singularities in the real emission diagrams of the perturbative series and can be resummed with the same techniques used for other large logarithms. Threshold resummation has been applied to several Standard Model processes, including (inclusive) Drell-Yan \cite{Sterman:1986aj, Catani:1989ne, Idilbi:2005ky,Becher:2007ty}, prompt photon \cite{Catani:1998tm,Becher:2009th}, Higgs production \cite{Moch:2005ky,Laenen:2005uz,Idilbi:2005ni,Ahrens:2008qu,Ahrens:2008nc}, and dijet and heavy-particle production \cite{Laenen:1991af,Berger:1995xz,Catani:1996yz,Kidonakis:1998bk,Kidonakis:1997gm,Kidonakis:1998nf}.

It is important to understand when threshold logarithms are phenomenologically relevant. The partonic endpoint (where logarithms of z become large) cannot be observed, since one integrates over the partonic variable $z$ when the partonic cross section is convolved with parton distribution functions (pdfs). The exception is near the hadronic endpoint, where the partonic and hadronic endpoints are equivalent. As an example, consider Drell-Yan, where the rescaled leptonic center of mass energy is given by $\tau = m^2_{\ell\ell}/s$. The hadronic cross section is then
\begin{equation}
\label{eq:DY_convolution}
\frac{\df \sigma}{\df \tau} = \int_\tau^1 \df z\ts {\cal L}(\tau/z) \hat \sigma(\tau,z)\,,
\end{equation}
where 
\begin{equation}
\label{eq:integrated_luminosity}
{\cal L}(z) = \int_{-\log\frac{z}{\tau}}^{\log\frac{z}{\tau}}\df Y\ts \left[\fone\ftwo+f_1\left(\xtwo\right)f_2\left(\xone\right)\right]
\end{equation}
is the parton luminosity. From this expression, one can immediately see that a measurement of Drell-Yan close to the hadronic endpoint ($\tau \to 1$) also forces the partonic center of mass energy $z$ to its endpoint, such that these logarithms become important. However, away from the hadronic endpoint, the integration variable $z$ is not forced to one, such that it is not clear that the threshold terms given in Eq.~(\ref{eq:threshold_terms}) should dominate over terms which are, for example, polynomial in $z$.

By comparing resummed cross sections with available fixed order calculations it has been noted that, in certain cases, the threshold terms at ${\cal O}(\alpha_s)$ give the dominant effect of the full NLO correction, even far away from the hadronic endpoint~\cite{Appell:1988ie,Catani:1998tm,Becher:2007ty}. This observation has been used to suggest that threshold resummation is effective not only in the hadronic endpoint, where it is clearly necessary, but also for much lower values of $\tau$. One possible explanation for this is the shape of parton luminosities~\cite{Appell:1988ie,Catani:1998tm}. For steeply falling parton luminosities, the integration in Eq.~(\ref{eq:DY_convolution}) is dominated by the smallest values of $\tau/z$, which is the region $z \to 1$. By assuming a simple, analytical form for the parton luminosity, one can make this argument more precise \cite{Becher:2007ty}. However, the corrections to threshold resummation applied away from hadronic endpoint have not yet been studied without assuming a model for parton luminosities.

The goal of this paper is to provide a model-independent, quantitative measure of the corrections to partonic threshold resummation. Using only the shape of the pdfs, we define a parameter $\lambda$ that can be used as the expansion parameter in SCET. This implies that corrections to the threshold resummation are power suppressed in $\lambda$. Since we will use the measured pdfs to calculate $\lambda$, our definition does not require the assumption of a particular form for the parton luminosities. Thus, $\lambda$ can be viewed as a model-independent definition of the steepness of a parton luminosity. We will determine the numerical size of $\lambda$ for various parton luminosities as a function of the hadronic threshold variables $\tau$ and $Y$. In addition to quantifying power corrections, $\lambda$ allows us to determine the scales relevant to threshold resummation in SCET after convolving the partonic cross section with the pdfs. This avoids integrating over unphysical regions (a consequence of setting the scales before integration) or introducing a process-specific procedure to calculate the scales.

This paper is organized as follows. In Section~\ref{sec:definition}, we will define the relevant expansion parameter in threshold resummation and show that it has the expected behavior for a simple choice of the pdfs. In Section~\ref{sec:numerical}, we will determine the numerical values of this expansion parameter for the actual pdfs, and present our conclusions in Section~\ref{sec:conclusions}. In Appendix~\ref{sec:consistency}, we show the consistency of the factorization theorem assuming only small values of $\lambda$, but keeping an arbitrary functional form for the pdfs, while in Appendix~\ref{sec:scaling} we discuss the scaling in $\lambda$ of integrals against the pdfs.

\section{Definition of Steepness}
\label{sec:definition}

The only assumptions one has to make to allow for the resummation of threshold logarithms is that $1-z \ll 1$, where $z$ is appropriately defined, along with the statement that pdfs factorize from the partonic cross section \cite{Collins:1981ta, Collins:1981uk}. We will use the definitions
\begin{equation}
z \equiv \frac{\hat{s}_{\rm min}}{\hat s}\,,\qquad \qquad \tau\equiv\frac{\hat s_{\rm min}}{s}\,,
\end{equation}
with
\begin{equation}
\label{eq:Mmin}
\hat s_{\rm min} = \Big(q+ \sum_i^N p_{J}^i \Big)^2\,.
\end{equation}
Here $q$ denotes the sum of the momenta of all non-strongly interacting particles, and  $p_J^i$ is the momentum of the $i$th jet. This momentum is defined in terms of $p_J^\T$ and $\eta_J$ of the jet as
\begin{equation}
\label{eq:masslessPJ}
p_J \equiv (p_J^\T \cosh \eta_J, \vec{p}_J^\T, p_J^\T \sinh \eta_J)
\,.\end{equation}
Thus, the jet 4-momentum is reconstructed from the measured transverse momentum and rapidity, assuming zero mass. 

A generic hadronic cross section can be written as 
\begin{equation}
\label{eq:sigma_hadronic}
\df \sigma = \int_{-\frac12 \ln \frac1\tau}^{\frac12 \ln \frac1\tau} \df Y \int_{\tau e^{-2|Y|}}^{1} \df z\ts 
\fone\ftwo\df \hat \sigma(z, \tau, Y)+(1\leftrightarrow 2)\,.
\end{equation}
One can use soft-collinear effective theory (SCET) to analyze the partonic cross section, with corrections being suppressed by powers of $1-z$
\begin{equation}
\label{eq:SCETcorrections}
\df \hat \sigma(z, \tau, Y) = \df \hat \sigma(z, \tau, Y)_{\rm SCET} + {\cal O}(1-z)\,.
\end{equation}
As was shown in~\cite{Bauer:2010vu}, the SCET partonic cross section factorizes, and renormalization group (RG) equations can be used to resum threshold logarithms. This work made no assumptions about the kinematics of the event other than $1-z\sim\lambda^2$, where $\lambda$ is the usual SCET power counting parameter, which relates the hard, collinear and soft scales via $\mu_c/\mu_h \sim \mu_s / \mu_c \sim \lambda$.

From Eq.~(\ref{eq:sigma_hadronic}) it is clear that $z$ is forced to one if the kinematics force us close to hadronic threshold, $\tau \to 1$. In this case, $1-z\ll 1$, which means that the power corrections in Eq.~(\ref{eq:SCETcorrections}) are small. Away from hadronic threshold, however, the partonic cross section gets integrated over values of $z$ for which $1-z \sim {\cal O}(1)$. For generic parton luminosities, threshold resummation will have ${\cal O}(1)$ power corrections when applied away from hadronic threshold, if the region $1-z\sim {\cal O}(1)$ is large.

If the parton luminosities are dominated by the region where $1-z \ll 1$, however, the SCET expression gives the dominant contribution to the hadronic cross section. We can quantify the corrections due to this by defining a parameter
\begin{equation}
\label{epsilondef}
\epsilon(\lambda,\tau,Y) = \frac{\int_{\tau e^{-2|Y|}}^{1-\lambda^2} \df z \, 
\left| \fone \right|\left| \ftwo \right|}{\int_{1-\lambda^2}^{1} \df z \, 
\left| \fone \right|\left| \ftwo \right|}\,.
\end{equation}
In other words, the parameter $\epsilon$ measures how much of the parton luminosity is contained in the region $\tau e^{-2|Y|} < z < 1-\lambda^2$ compared to the region $1 > z > 1-\lambda^2$. This parameter obviously depends not only on our choice of $\lambda$, but also on the hadronic center of mass energy $\tau$ and the total rapidity of the event $Y$. Note that for fixed values of $\tau$ and $Y$, $\epsilon$ is as a monotonically decreasing function of $\lambda$. 

Using this definition, we can write
\begin{equation}
\label{epsilon_lambda_power_corrections}
\df \sigma = \int_{-\frac12 \ln \frac1\tau}^{\frac12 \ln \frac1\tau} \df Y \int_{1-\lambda^2}^{1} \df z \, 
\fone\ftwo\df \hat \sigma(z, \tau, Y)_{\rm SCET} + 
{\cal O}(\epsilon, \lambda^2)\,,
\end{equation}
where the ${\cal O}(\lambda^2)$ correction comes from the fact that $1-z$ in Eq.~(\ref{epsilon_lambda_power_corrections}) is bounded by $\lambda^2$. The total correction to threshold resummation can therefore be estimated as ${\rm max} (\epsilon, \lambda^2)$. However, since $\epsilon$ decreases with increasing $\lambda^2$, the total correction is minimized if we choose $\epsilon = \lambda^2$. The corrections are then ${\cal O}(\lambda^2)$. This also allows us to define $\lambda^2$ through the relation
\begin{equation}
\label{lambdadefdouble}
\lambda^2 = \frac{\int_{\tau e^{-2|Y|}}^{1-\lambda^2} \df z \, 
\left| \fone \right|\left| \ftwo \right|}{\int_{1-\lambda^2}^{1} \df z \, 
\left| \fone \right|\left| \ftwo \right|}\,,
\end{equation}
where $\lambda^2\equiv\lambda^2(\tau,Y)$. We can also define $\lambda^2$ for a single pdf
\begin{equation}
\label{lambdadefsingle}
\lambda^2=\frac{
\int_x^{1-\lambda^2}\df z\ts \left|f\left(\frac{x}{z}\right)\right|}{
\int_{1-\lambda^2}^1\df z\ts \left|f\left(\frac{x}{z}\right)\right|}\,,
\end{equation}
where here $\lambda^2\equiv\lambda^2(x)$. This definition will be useful for determining if endpoint Altarelli-Parisi functions are valid when evolving the pdfs.

\begin{figure}[htbp]
\includegraphics[width=0.9\columnwidth]{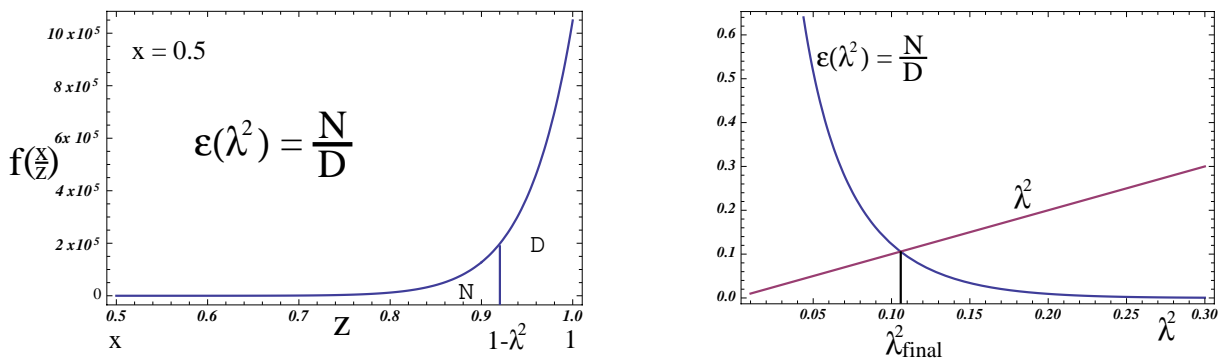}
\caption{Left: The parameter $\epsilon$, defined in Eq.~(\ref{epsilondef}), illustrated as the ratio of two areas. Right: For power corrections that go as max$(\epsilon,\lambda^2)$, there is a choice of $\lambda^2=\lambda^2_{\rm final}$ such that the corrections are minimized. This choice, called $\lambda^2$ in the text, is defined in Eqs.~(\ref{lambdadefdouble}) and (\ref{lambdadefsingle}).}
\label{fig:explanation}
\end{figure}
We illustrate our definitions in Fig.~\ref{fig:explanation}, with a simple example function $f(x)$. The graphic on the left shows $\epsilon$ as the ratio of two areas: the region D, where $1-z<\lambda^2$ and the region N, where $1-z>\lambda^2$. For $N\ll D$ and $\lambda^2\ll 1$, we see that the area under the curve is dominated by a region where $1-z\ll 1$. The right graphic shows that Eq.~(\ref{lambdadefsingle}) has a solution where power corrections are minimized, labelled $\lambda^2_{\rm final}$.

Independent of the form of the pdfs, we can see that $\lambda^2_{\rm final}$ goes to 0 as $\tau$ (or $x$) goes to 1, as expected. First, note that $\tau\to1$ forces $Y\to0$, which sets the lower bound of the integral in the numerator of Eq.~(\ref{epsilondef}) to $\tau$. Next, we can examine the limits on $\lambda^2$. As $\lambda^2$ goes to 0, $\epsilon$ goes to $\infty$ (here we've assumed that the pdf implicitly contains a $\Theta$ function that sets it to 0 if the argument is greater than 1 or less than 0). Similarly, as $\lambda^2$ goes to $1-\tau$, $\epsilon$ goes to 0. Therefore, the crossover point, where $\epsilon=\lambda^2$, is between 0 and $1-\tau$, which means $\lambda^2_{\rm final}\to0$ as $\tau\to1$. For the remainder of the paper, we will omit the subscript final on $\lambda^2$.

Note that these definitions do not assume anything about the form of the pdfs and are therefore completely model independent. Given that pdfs are not observable quantities, many pdf sets do not constrain them to be positive. For this reason, we have used absolute values of the pdfs in our definition. Of course, the value of $\lambda^2$ depends on the functional form of the pdfs, but this can be determined numerically using any of the available pdf sets. We will present the numerical results for the power counting parameter $\lambda^2$ in Section~\ref{sec:numerical}. 

We end this section by showing the behavior of $\lambda^2$ for a particularly simple form of the pdfs. This will illustrate our proposal with an analytic example and should convince the reader that corrections to threshold resummation behave as expected. The assumed form of the pdfs does not represent their observed shape and is only chosen so that the analytic results can be easily studied. All numerical results presented in Section~\ref{sec:numerical} will use the measured pdfs. 

Consider the simple model 
\begin{equation}
f_i(x) = x^{-a_i}\,,
\end{equation}
where the pdf becomes steeper as $a$ increases. 
Given this form, Eq.~(\ref{lambdadefdouble}) becomes
\begin{equation}
\lambda^2 = \frac{(1-\lambda^2)^{1+(a_1+a_2)/2} -(\tau e^{-2|Y|})^{1+(a_1+a_2)/2}}{1-(1-\lambda^2)^{1+(a_1+a_2)/2}}\,.
\end{equation}
To simplify this equation, we define
\begin{equation}
r \equiv A \lambda^2\,, \qquad {\rm with} \qquad A \equiv 1+\frac{a_1+a_2}{2}\,.
\end{equation}
The equation can then be written as
\begin{equation}
\left(1+\frac{r}{A}\right)\left(1-\frac{r}{A}\right)^A = \frac{r}{A} + (\tau e^{-2|Y|})^A\,.
\end{equation}
We can clearly see that $r \to 0$ in the hadronic endpoint $\tau \to 1$ (for which $Y \to 0$), independent of the value of the parameter $A$. However, for large values of $A$, and therefore steep pdfs, we can use the approximations $1+r/A \approx 1$ and  $(1-r/A)^A \approx e^{-r}$ to find
\begin{equation}
r = W(A)\,,
\end{equation}
where $W(A)$ denotes the product logarithm of $A$. Since the $W(A)$ only grows logarithmically with $A$, we find the desired result that $\lambda^2 \to 0$ as $A \to \infty$.

\section{Numerical results for $\lambda^2$}
\label{sec:numerical}

Now that we have defined a measure of steepness, it is a simple matter to determine numerically what $\lambda^2$ is for the physical pdfs and parton luminosities. In this section, we will give our results and discuss their implications for the relevance of threshold resummation. As discussed above, the expansion parameter for threshold resummation is given by $\lambda^2$, and we make the assumption that this expansion becomes reliable for $\lambda^2 < 0.25$. Note that while this value is chosen somewhat ad-hoc, it serves to assess the importance of threshold resummation. 

\begin{figure}[htbp]
\includegraphics[width=0.9\columnwidth]{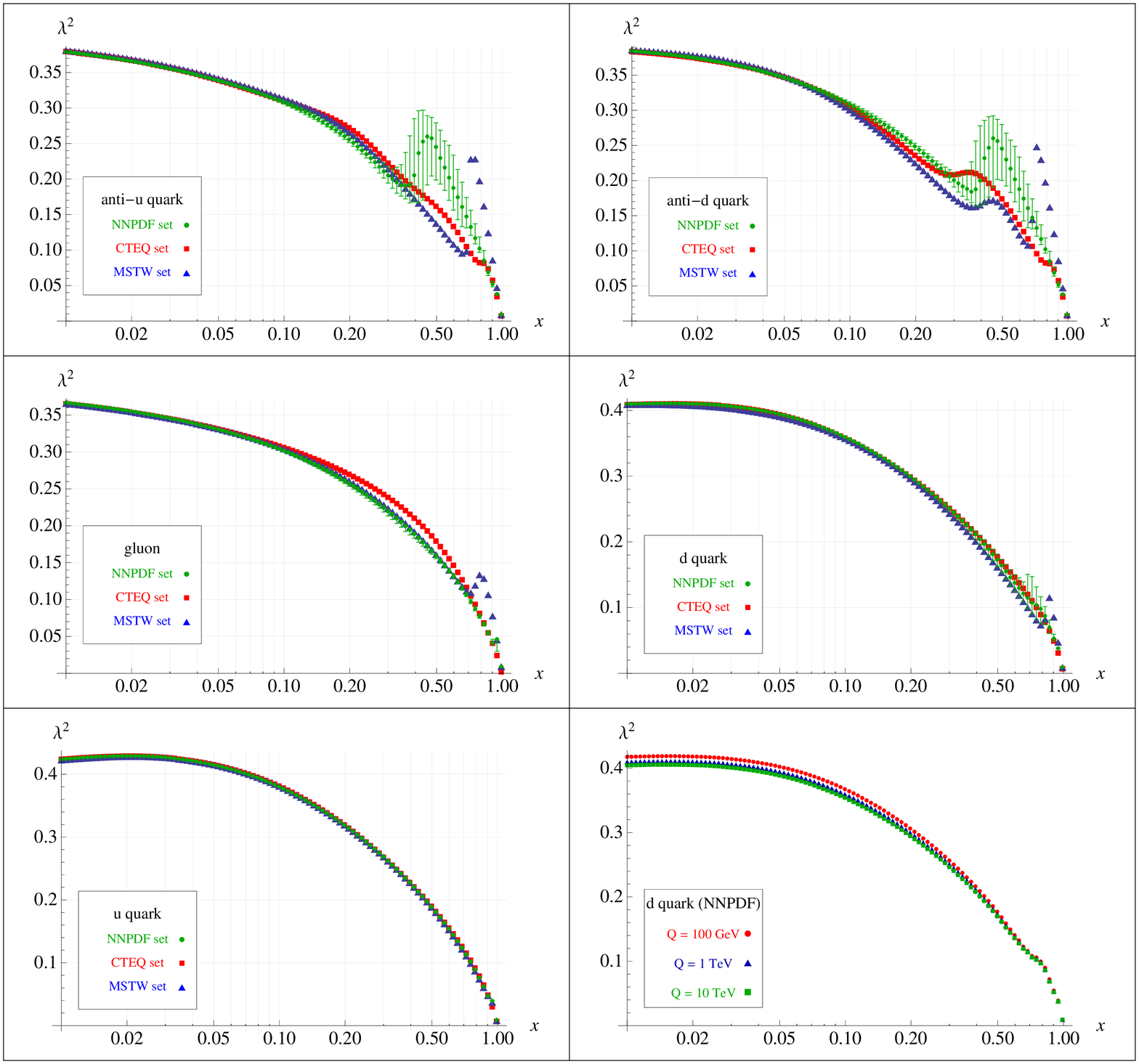}
\caption{The value of $\lambda^2$ for single pdfs as a function of $x$, as defined in Eq.~(\ref{lambdadefsingle}).}
\label{fig:singlepdf}
\end{figure}
\begin{figure}[htbp]
\includegraphics[width=0.9\columnwidth]{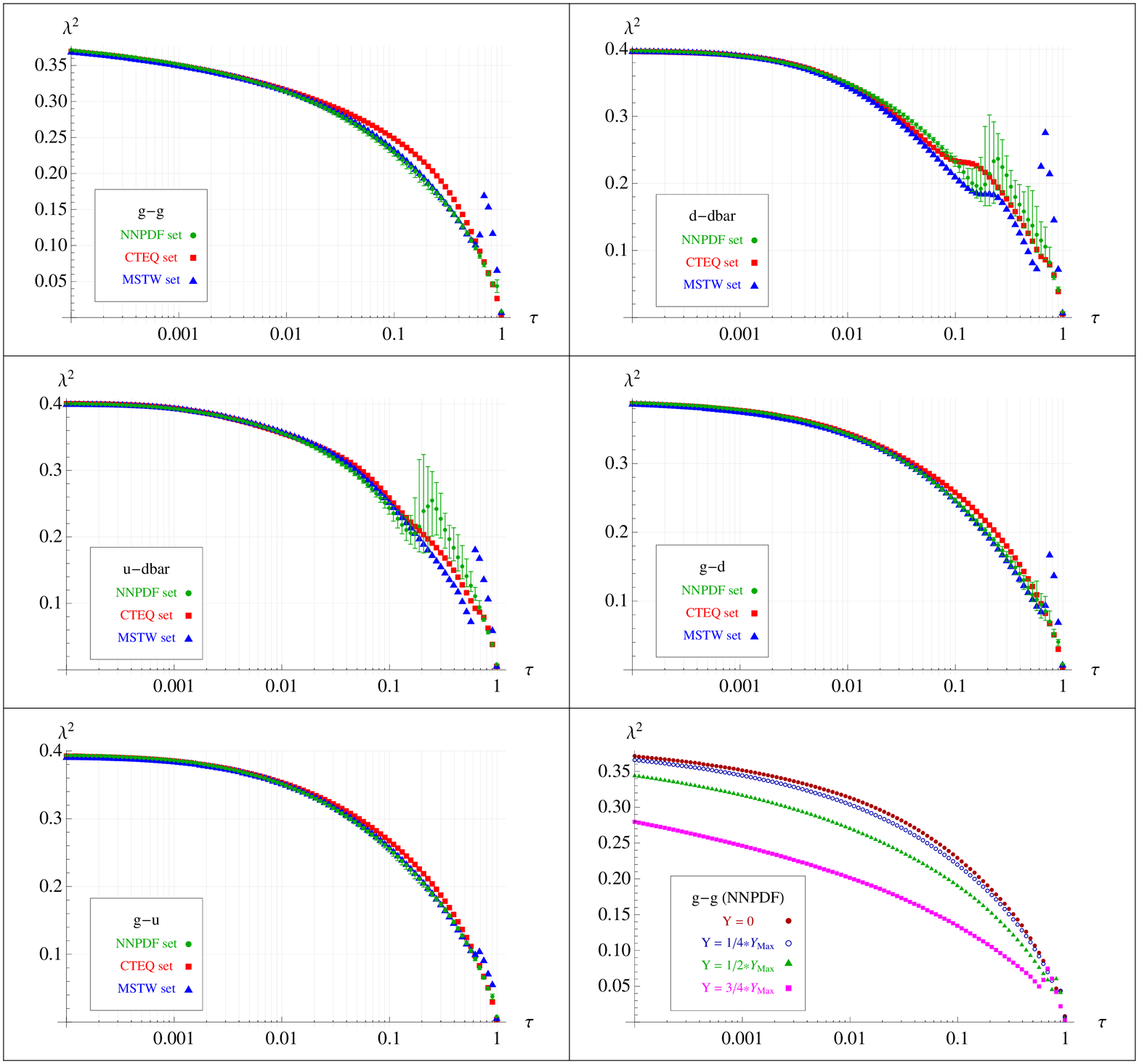}
\caption{The value of $\lambda^2$ for a selection of parton luminosities as a function of $\tau$, as defined in Eq.~(\ref{lambdadefdouble}).}
\label{fig:doublepdf}
\end{figure}

There are two different types of objects for which steepness is a necessary ingredient of our factorization theorem. The first is individual pdfs, where $\lambda^2 \ll 1$ allows us to use endpoint AP kernels in determining the consistency of our factorization theorem. The second is parton luminosities, in the form of two pdfs which are convolved in $Y$ and $z$. Since it is possible to be differential in $Y$ up to power corrections in $\lambda^2$ (see Appendix~\ref{sec:consistency}), we will focus on the case where $Y$ is fixed and $z$ is the only convolution variable. We will present plots using the CTEQ6.6~\cite{Pumplin:2002vw}, MSTW2008NLO~\cite{Martin:2009iq} and NNPDF2.0~\cite{Ball:2010de} pdf sets, with a default renormalization scale of 1 TeV. To study the uncertainties in our predictions for $\lambda$, we use the 100 replica set provided by NNPDF2.0. The central value is given by the median of the results, while the error bars show the range where 68\% of the points lie, with equal number of points on either side of the median. 

In Fig.~\ref{fig:singlepdf} we present the value of $\lambda^2$ for a single pdfs as a function of $x$, as defined in Eq.~(\ref{lambdadefsingle}), which characterizes how important the non-singular terms in the AP evolution kernels are. In each plot we show the result for CTEQ, MSTW and NNPDF sets, with errors presented for the NNPDF set. One clearly sees that the value of $\lambda^2$ is decreasing as $x \to 1$, as expected, and that $\lambda^2$ is small not only in the region where $1-x \ll 1$ but for smaller values of $x$ as well.  Irrespective of the type of parton, the value of $\lambda$ becomes smaller than $0.25$ for $x\gtrsim 0.3$, implying that in that region the singular AP evolution kernel becomes a reasonable approximation to QCD, even though $1-x = {\cal O}(1)$. The feature in the sea-quark distributions for $x \sim 0.5$ is due to the steepness falling off in the measured pdfs, giving a rising value of $\lambda^2$, which then gets forced to zero due to $x$ approaching 1. In the lower right plot of Fig.~\ref{fig:singlepdf}, we illustrate the dependence of steepness on the renormalization scale of the pdfs. Between 100 GeV and 10 TeV, there is at most an ${\cal O} (5\%)$ variation, while for most values of x the variation is ${\cal O} (1\%)$. In this plot, we only present the results for the down quark, however, the effect is similar (${\cal O}(1-5\%)$) for all other pdfs. 

Next, we study the value of $\lambda^2$ as obtained from the convolution of two pdfs, as defined in Eq.~(\ref{lambdadefdouble}). The results are shown in Fig.~\ref{fig:doublepdf} for $Y = 0$. While the distributions in general do not peak exactly at $Y=0$, the peak is sufficiently close to this value for these plots to illustrate the observed behavior. Again we show the result for CTEQ, MSTW and NNPDF sets, with errors presented for the NNPDF set. The value of $\lambda^2$ becomes less that 0.25 for $\tau \gtrsim 0.05-0.1$. This implies that threshold resummation is not limited to the region where $1-\tau \ll 1$, however, for most values of phenomenological interest, the power corrections are potentially large. In the lower right hand side of  Fig.~\ref{fig:doublepdf}  we show the variation of steepness with $Y$, presented as a function of $Y_{\rm max} = -(1/2)\,  \ln \tau$. While this variation is considerable, one should keep in mind that the majority of the phase space is close to $Y=0$. The variation of $\lambda^2$ with renormalization scale is not shown for the parton luminosities, but is similar to the result for the single pdfs.

\section{Conclusions}
\label{sec:conclusions}
In this paper we presented a model-independent, quantitative measure of the power corrections to partonic threshold resummation. Previously, any quantitative study of the power corrections associated with applying hadronic threshold resummation away from the true endpoint necessitated simplified assumptions about the functional form of the pdfs. The expansion parameter of threshold resummation is given by a parameter $\lambda^2$, which can be defined unambiguously through integrals over pdfs. In SCET, the parameter $\lambda^2$ is related to the scale of soft and collinear radiation, such that the scales $\mu_c$ and $\mu_s$ can be chosen after convolving the partonic cross section with the pdfs. This allows us to avoid integrating through regions where the perturbative expansion is ill-defined, without having to define an arbitrary procedure for determining scales after integration. We have shown analytically that the value of $\lambda$ approaches zero either in the true hadronic endpoint or in the limit of infinitely steep pdfs. Our numerical results indicate that while threshold resummation can be justified away from the region where $1-\tau \ll 1$, we expect the corrections to become sizable for $\tau \lesssim 0.05-0.1$. 
\begin{acknowledgments}
This work was supported by the Director, Office of Science, Offices of High Energy and Nuclear Physics of the U.S. Department of Energy under the Contracts DE-AC02-05CH11231. AH was in addition supported by a LHC Theory Initiative Fellowship with NSF grant number PHY-0705682, and the DOE under contract DE-FGO3-96-ER40956.
\end{acknowledgments}

{\it Note that while this paper was being prepared for publication, a paper with somewhat overlapping contents appeared~\cite{Bonvini:2010tp}.}

\begin{appendix}

\section{Consistency of threshold resummation}
\label{sec:consistency}

As discussed in the introduction, threshold resummation can be derived using effective field theory methods. Using SCET, and assuming $1-z \sim \lambda^2 \ll 1$ as well as $\df\sigma = f\otimes f\otimes\df\hat\sigma$, one can derive a factorization theorem for a generic differential cross section. We refer the reader to~\cite{Bauer:2010vu} for the details of this derivation. One check of this factorization theorem is consistency, which is the $\mu$ independence of the factorized result. In~\cite{Bauer:2010vu} we showed the consistency of our factorization theorem using the partonic definition of the pdfs $f_i(x;\mu) = \delta(1-x)$. Since we have just shown that away from the true hadronic endpoint it is the functional form of the pdfs that defines the expansion parameter $\lambda$, we wish to repeat this proof using the full functional form, subject only to the constraint of large steepness (or small $\lambda$). 

We start from the derivative with respect to $\ln\mu$ of the factorized differential cross section derived in~\cite{Bauer:2010vu}, together with the anomalous dimensions for the hard, jet, and soft functions, given in the same reference. Using the result of Appendix~\ref{sec:scaling} for the Altarelli-Parisi splitting kernel, and performing trivial integrations and cancellation, we find
\begin{align}
\label{eq:preintegral}
\dlog{\sigma}&\propto\int\df Yf_1(\sqrt{\tau}e^Y)f_2(\sqrt{\tau}e^{-Y})\Biggl[
\vec{T}_1^2\ts\ln\frac{\omega_1}{\sqrt{\hat{s}}}+\vec{T}_2^2\ts\ln\frac{\omega_2}{\sqrt{\hat{s}}}+\sum_{\ijpair}\vec{T}_i\mcdot\vec{T}_j\ts\ln\frac{\tilde{n}_i\mcdot\tilde{n}_j}{n_i\mcdot n_j}+\sum_{i\in\jetset}\vec{T}_i^2\ts\ln\frac{\omega_i}{\tilde{\omega}_i}
\new
&
+ \vec{T}_1^2\Biggl(\int_{\sqrt{\tau}e^Y}^1\df w_1\frac{f_1\left(\frac{\sqrt{\tau}e^Y}{w_1}\right)}{f_1(\sqrt{\tau}e^Y)}\frac{1}{(1-w_1)_+}
-\int_{\tau e^{\abs{2Y}}}^1\df z\ts\frac{f_1\left(\sqrt{\frac{\tau}{z}}e^Y\right)f_2\left(\sqrt{\frac{\tau}{z}}e^{-Y}\right)}{f_1\left(\sqrt{\tau}e^Y\right)f_2\left(\sqrt{\tau}e^{-Y}\right)}\frac{1}{(1-z)_+}\Biggr)
\new
&
+\vec{T}_2^2\Biggl(\int_{\sqrt{\tau}e^{-Y}}^1\df w_2\frac{f_2\left(\frac{\sqrt{\tau}e^{-Y}}{w_2}\right)}{f_2(\sqrt{\tau}e^{-Y})}\frac{1}{(1-w_2)_+}
-\int_{\tau e^{\abs{2Y}}}^1\df z\ts\frac{f_1\left(\sqrt{\frac{\tau}{z}}e^Y\right)f_2\left(\sqrt{\frac{\tau}{z}}e^{-Y}\right)}{f_1\left(\sqrt{\tau}e^Y\right)f_2\left(\sqrt{\tau}e^{-Y}\right)}\frac{1}{(1-z)_+}\Biggr)
\Biggr]\,.
\end{align}
The first line in Eq.~(\ref{eq:preintegral}) can be shown to vanish using the identity
\begin{equation}
\sum_{\ijpair}\vec{T}_i\mcdot\vec{T}_j\ts\ln\frac{\tilde{n}_i\mcdot\tilde{n}_j}{n_i\mcdot n_j}=\sum_{i\in\partset}\vec{T}_i^2\ts\ln\frac{\tilde{\omega}_i}{\omega_i}\,.
\end{equation}
The first term in the second line can be made identical to the second term through the substitution $Y^\prime = Y - \frac{1}{2}\log w_1$, followed by the relabeling $Y^\prime\to Y$. Thus these two terms cancel, and using a similar substitution $Y^\prime = Y + \frac{1}{2}\log w_2$, with the same relabeling, the third line vanishes as well. Thus, we find
\begin{equation}
\dlog{\sigma}=0+{\cal O}(\lambda^2)\,,
\end{equation}
and therefore consistency of the factorization formula. 
Note that in both cases $Y^\prime=Y+\order{\lambda^2}$, which means we can be differential in $Y$, up to power corrections in $\lambda^2$, and still have consistency.

\section{General power corrections}
\label{sec:scaling}
It is instructive to understand the power corrections and scaling of various test functions integrated against a steep distribution, $f(z)$, where steep is defined as
\begin{equation}
\frac{
\int_{\zmin}^{1-\lambda^2}\df z\ts f(z)}{
\int_{1-\lambda^2}^1\df z\ts f(z)}
= \lambda^2\ll 1\,.
\end{equation}
We define all scaling relative to $\int_{1-\lambda^2}^1\df z\ts f(z)$, which can be taken to be ${\cal O}(1)$ without loss of generality.

The simplest choice for a test function is a non-singular polynomial of $z$, $g(z)$. In general, we can express $g$ as a Taylor series about $z=1$,
\begin{equation}
g(z)=\sum_{n=0}^{\infty}\frac{g^{(n)}(1)}{n!}(z-1)^n\,,
\end{equation}
where we assume all $g^{(n)}(1)$ are ${\cal O}(1)$. We can evaluate the integral of $f(z)g(z)$ by dividing the integration region into two parts,
\begin{equation}
\int_{\zmin}^1\df z\ts f(z)g(z) = \int_{\zmin}^{1-\lambda^2}\df z\ts f(z)g(z)+\int_{1-\lambda^2}^1\df z\ts f(z)g(z)\,.
\end{equation}
In the region $\zmin < z < 1-\lambda^2$, the integral is $\order{\lambda^2}$, since $f(z)$ and $f(z)g(z)$ are of the same order and the integral of $f(z)$ over this region is $\order{\lambda^2}$. For $1-\lambda^2 < z < 1$,
\begin{align}
\int_{1-\lambda^2}^1\df z\ts f(z)g(z) &= \sum_{n=0}^\infty\frac{g^{(n)}(1)}{n!}\int_{1-\lambda^2}^1\df z\ts f(z)(z-1)^n\new
&\sim\sum_{n=0}^\infty\frac{g^{(n)}(1)}{n!}\int_{1-\lambda^2}^1\df z\ts f(z)\times\left(\lambda^2\right)^n\,.
\end{align}
In this region, the integral is dominated by the zeroth order term in the power expansion. Therefore, we can write the integral of $g(z)$ against the distribution $f(z)$ as
\begin{equation}
\int_{\zmin}^1\df z\ts f(z)g(z) = g(1)\int_{1-\lambda^2}^1\df z\ts f(z)+\order{\lambda^2}\,.
\end{equation}

A second possible test function is a $\delta$-function at $z=1$. The integral is trivially $f(1)$, but we would like to know the scaling of this result. Assuming $f$ is monotonically increasing from $\zmin$ to 1,
\begin{equation}
\int_{1-\lambda^2}^1\df z\ts f(z)\leq f(1)\times\lambda^2\,.
\end{equation}
Since this integral is $\order 1$, we can say that $f(1)$ is $\order{\lambda^{-2}}$.

The last class of test functions we will consider is $\log^n(1-z)/(1-z)$ plus functions, multiplied by a non-singular function $g(z)$. When integrated against $f(z)$, this gives
\begin{align}
\label{eq:plus_power}
\int_{\zmin}^1&\df z\ts f(z)g(z)\left(\frac{\log^n(1-z)}{1-z}\right)_+\new
=&\int_{\zmin}^1\df z\ts\log^n(1-z)\frac{f(z)g(z)-f(1)g(1)}{1-z}-f(1)g(1)\int_0^{\zmin}\df z\ts\frac{\log^n(1-z)}{1-z}\new
=&\int_{1-\lambda^2}^1\df z\ts\log^n(1-z)\frac{f(z)g(z)-f(1)g(1)}{1-z}+\int_{\zmin}^{1-\lambda^2}\df z\ts\log^n(1-z)\frac{f(z)g(z)}{1-z}+f(1)g(1)\frac{\log^{n+1}\lambda^2}{n+1}\,.
\end{align}
Since $f(z)$ is at most $\order{\lambda^2}$ over the interval $\zmin < z < 1-\lambda^2$, the second term is clearly power suppressed relative to the final term. Focusing on the first term, if we express $g$ as a Taylor series about $z=1$, we see that the zeroth order term is again dominant. This gives
\begin{equation}
\int_{\zmin}^1\df z\ts f(z)g(z)\left(\frac{\log^n(1-z)}{1-z}\right)_+=g(1)\int_{1-\lambda^2}^1\df z\ts f(z)\left(\frac{\log^n(1-z)}{1-z}\right)_++\order{\lambda^2}\,.
\end{equation}
Note that Eq.~(\ref{eq:plus_power}) shows that the scaling of this integral is $\order{\lambda^{-2}\log^{n+1}\lambda^2}$ and therefore, for small $\lambda^2$, threshold effects are important.

The Altarelli-Parisi (AP) kernels, $P_{ij}(z)$, are polynomial functions of $z$ for $i\neq j$ and a combination of $\delta$-functions and plus functions for $i = j$. This means that for steep pdfs (assuming all $\lambda_i^2$ are small), the $P_{ij}$ for $i\neq j$ are suppressed in $\lambda^2$ compared to $P_{ii}$. Moreover, any polynomial functions of $z$ in $P_{ii}(z)$ can be evaluated at $z = 1$, up to power corrections in $\lambda^2$. The result is endpoint AP kernels, given at NLL accuracy by
\begin{equation}
\frac{\as}{\pi}P_{ij}(z) = \left[2 \ts\Gamma_{\cusp}\ts\vec{T}_i^2\left(\frac{1}{1-z}\right)_++\gamma_i\ts\delta(1-z)\right] \delta_{ij}\,.
\end{equation}

\end{appendix}



\bibliography{bibliography}

\end{document}